\documentclass[10pt,letterpaper]{article}

\usepackage{opex3}

\usepackage{amsmath}
\usepackage[verbose]{cite}

\allowdisplaybreaks[1]

\begin{document}

%%%%%%%%%%%%%%%%%% title page information %%%%%%%%%%%%%%%%%%
\title{An analytical investigation of optical vortices emitted from a collectively polarized dipole array}

\author{Motoki Asano$^{1*}$ and Tohru Takahashi$^{2}$}

\address{$^1$ Graduate School of Engineering Science, Osaka University, Toyonaka, Osaka 560-8531, Japan \\
$^2$Department of Electrical Engineering, Oita National College of Technology, Maki, Oita, 870-01, Japan}

\email{$^*$asano@qi.mp.es.osaka-u.ac.jp} %% email address is required


\begin{abstract}
Many approaches for producing optical vortices have been developed both for fundamental interests of science and for engineering applications. In particular, the approach with direct excitation of several emitters has a potential to control the topological charges with a control of the source conditions without any modifications of structures of the system. In this paper, we investigate the propagation properties of the optical vortices emitted from a collectively polarized electric dipole array as a simple model of the several emitters. Using an analytical approach based on the Jacobi-Anger expansion, we derive a relationship between the topological charge of the optical vortices and the source conditions of the emitter, and clarify and report our new finding; there exists an intrinsic split of the singular points in the electric field due to the spin-orbit interaction of the dipole fields.
\end{abstract}

\ocis{(050.4865) Optical vortices; (260.6042) Singular Optics.} 

%%%%%%%%%%%%%%%%%%%%%%% References %%%%%%%%%%%%%%%%%%%%%%%%%

%%%%%%%%%%%%%%%%%%%%%%%%%%  body  %%%%%%%%%%%%%%%%%%%%%%%%%%
\section{Introduction}

Optical vortices, which carry orbital angular momentum of light, are characterized by a spatially distributed phase profile of $e^{il\phi}$ where $\phi$ is an azimuthal angle on the plane orthogonal to the propagation direction of light, and $l$ is the topological charge of optical vortices \cite{allen1992orbital,molina2007twisted, yao2011orbital}. Optical vortices has been actively studied over a wide range of fields such as optical manipulation \cite{grier2003revolution}, optical sensing \cite{lavery2013detection}, laser fabrication\cite{omatsu2010metal} and classical and quantum information processing \cite{wang2012terabit,mair2001entanglement}. In order to generate and control the optical vortices, the mode conversion of electromagnetic field has been mainly proposed in various structural devices. The spatial light modulators were used to generate a Laguerre-Gaussian beam with an arbitrary topological charge by holographic patterns \cite{heckenberg1992generation, matsumoto2008generation}. The pinhole plates with various arrangements were designed to create arrays of optical vortices by illuminating with a plane wave \cite{nicholls1987three,li2013generation}. The spin-to-orbital angular momentum conversion were performed using artificial optical devices such as Q-plate \cite{marrucci2006optical} and a plasmonic metasurface \cite{karimi2014generating}. The compact optical vortex emitters were fabricated to convert the optical modes from whispering-gallery-modes to optical vortices \cite{savchenkov2006optical,cai2012integrated}. On the other hand, the direct excitation of several emitters for producing optical vortices were recently studied in a ring molecules\cite{williams2014direct} and an antenna array \cite{ma2015optical}. This method has a potential to produce optical vortices by controlling the source conditions of the emitters without any modification of the system structure. 

In this paper, we analytically investigate the propagation properties of the optical vortices generated by a collectively polarized electric dipole array, which becomes a simple model for the direct excitation of several emitters. We derive an analytical solution of the electromagnetic field decomposed by vortex fields via several approximations, and clarify the following properties: (1) there is a certain relationship between the topological charge of the optical vortices and the source conditions of the emitters, and (2) There exists an intrinsic split of the singular points in the electric field due to the spin-orbit interaction of light between the collective electric dipole moments and the intrinsic radiation patterns of each emitter without any split of the singular point in the magnetic field.

\section{Theory}
\subsection{Far-field electromagnetic field decomposed by Jacobi-Anger expansion}
In our system, $N$ infinitesimal electric dipoles are placed on the x-y plane with the radius of $a$ such that the center of the circle is located at the origin of the coordinate. The coordinate of the $j$-th dipole is given by $(a\cos\varphi_j,a\sin\varphi_j,0)$ where the azimuthal angle $\varphi_j=\frac{2\pi (j-1)}{N}$. The electric dipole moment of all of the electric dipoles is represented by a collective polarization vector ${\bf p}=p(s_x {\bf e}_x+s_y{\bf e_y})e^{iq\varphi_j}$ where the real value $p$ shows the amplitude of the electric dipole moment, the complex parameters of $s_x, s_y$ represent the polarization of each dipole with the normalized amplitude $\sqrt{s_x^2+s_y^2}=1$ and $e^{iq\varphi_j}$ is the phase factor of each electric dipole with an integer $q$ which decides the topological property of the initial phase distribution of the system. The source condition of the system is characterized by the three parameters; the number of dipoles in the array $N$, the polarization of each dipole $(s_x,s_y)$ and the initial topological factor $q$. Figure 1(a) depicts an illustration of our system with the source condition of $(s_x,s_y)=(1,0)$, $N=3$ and $q=1$. An analytical solution of the electromagnetic field generated by the electric dipole array is given by the superposition of each electric dipole field \cite{jackson_classical_1999},
\begin{align}
&{\bf E}({\bf r})=\sum_{j=1}^N\frac{e^{ik\left|\pmb{ \xi}_j\right|}}{4\pi\varepsilon_0}\Biggl\{k^2\left(\frac{\pmb{ \xi}_j}{\left|\pmb{ \xi}_j\right|}\times{\bf p}\right)\times\frac{\pmb{ \xi}_j}{\left|\pmb{ \xi}_j\right|^2}+\left[3\frac{\pmb{ \xi}_j}{\left|\pmb{ \xi}_j\right|}\left(\frac{\pmb{ \xi}_j}{\left|\pmb{ \xi}_j\right|}\cdot{\bf p}\right)-{\bf p}\right]\left(\frac{1}{\left|\pmb{ \xi}_j\right|^3}-\frac{ik}{\left|\pmb{ \xi}_j\right|^2}\right)\Biggr\} \label{1}\\
&{\bf H}({\bf r})=\sum_{j=1}^N\frac{ck^2e^{ik\left|\pmb{ \xi}_j\right|}}{4\pi}\left(\frac{\pmb{ \xi}_j}{\left|\pmb{ \xi}_j\right|}\times {\bf p}\right)\frac{1}{\left|\pmb{ \xi}_j\right|}\cdot\left(1-\frac{1}{ik\left|\pmb{ \xi}_j\right|}\right)\label{2}
\end{align}
where $\pmb{ \xi}_j\equiv{\bf r}-{\bf R}_j$, ${\bf R}_j$ is the coordinate vector of the $j$th electric dipoles, $k$ is the magnitude of wavevector, $\varepsilon_0$ is the dielectric constant in the vacuum and the terms of the harmonic time evolution $e^{i\omega t}$ is omitted here. Because the profile of the electromagnetic field emitted from the dipole array is symmetrical about the $xy$ plane including the origin, we focus on the optical vortices propagating to the positive direction of the $z$ axis in the following discussion.

In order to verify the possibility for the generation of the optical vortices in our system, we calculate the phase distribution of the electromagnetic field in the case where the electric dipoles are collectively polarized to $x$ direction ($s_x=1,s_y=0$). Figure 1(b)--1(g) shows the phase distribution of the $x$ component of the electric and the $y$ component of the magnetic field with the three different couple of the parameters $(N,q)=(3,1),(5,2)$ and $(7,3)$ at $z=100\lambda$. The horizontal and the vertical axes are scaled by the wavelength of the electromagnetic field emitted from each dipole $\lambda$. The magnetic field has a singular point at the center of the $xy$ plane with a topological charge $q$. On the other hand, the singular points of the electric field appears with the split in the case of the higher order of $q$. The distance between the singular points is as long as the wavelength of the electromagnetic field. Although the exact solutions of Eq. (\ref{1}) and Eq. (\ref{2}) give us a certain distribution of electromagnetic filed, however, it is not clear for the relationship between the propagation properties of the optical vortices and the source conditions.
\begin{figure}[htbp]
\centering\includegraphics[width=13cm]{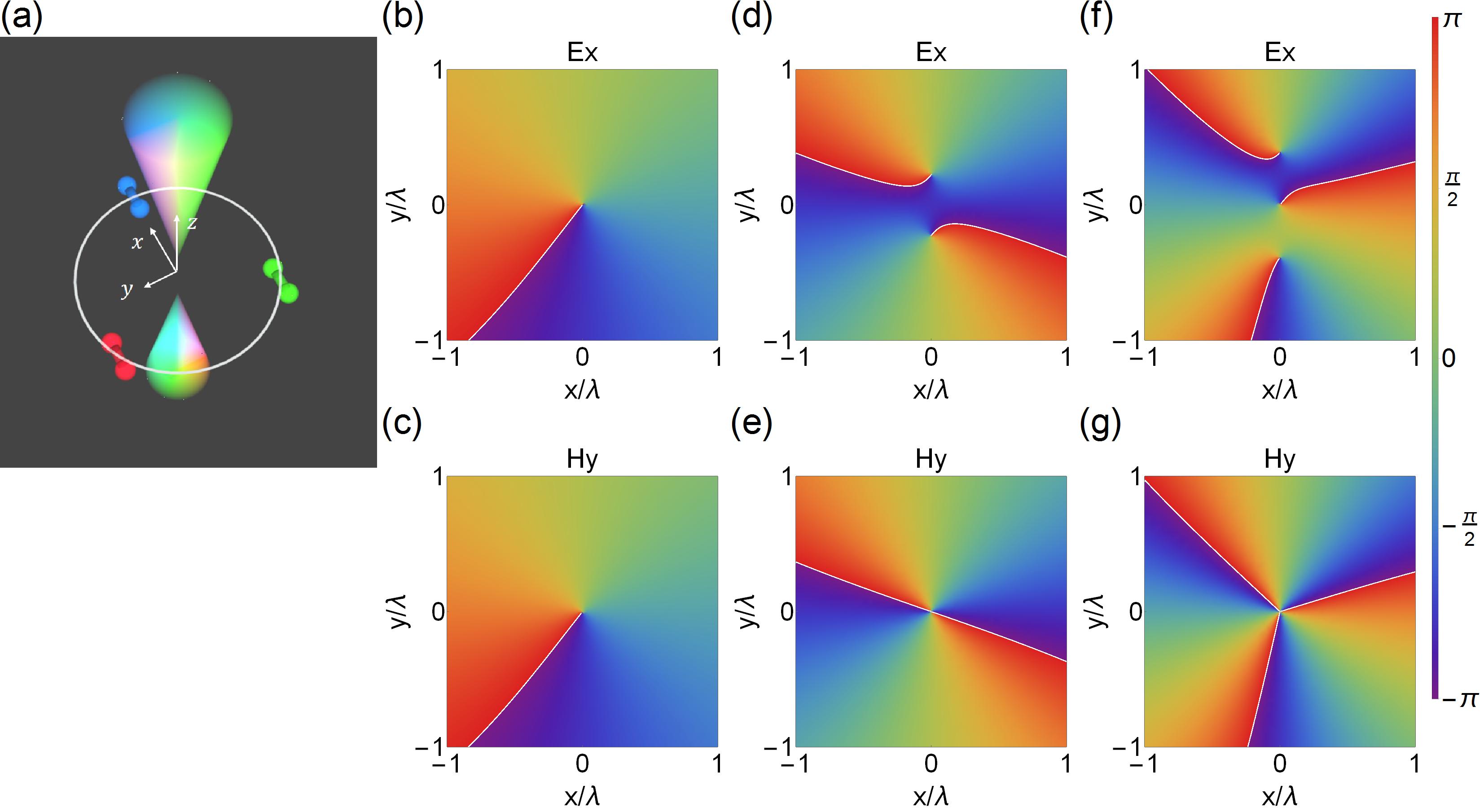}
\vspace{10pt}
\caption{Optical vortices of $x$ component of electric field generated by an electric dipole array which is collectively polarized to $x$ direction at $z=100\lambda$. (a) An illustration of our system.(b),(d) and (f) Phase distribution of electric field in the case of $(N.q)=(3,1)$, $(5,2)$ and $(7,3)$ respectively. (c), (e) and (f) Phase distribution of magnetic field in the case of $(N.q)=(3,1)$, $(5,2)$ and $(7,3)$ respectively.}
\end{figure}

In order to discuss the singular points in the electric fields split in the far-field with the distance as long as the wavelength of $\lambda$, we perform two approximations: the far-field approximation in which the higher terms of $\lambda/z\sim a/z$ are neglected and the near-axis approximation in which the higher terms of $\lambda/z\sim x/z\sim y/z$ are neglected. These two approximations are suitable for analyzing the internal phase structure of the optical vortices in the far-field regime. At first, we perform the approximation to the function of the phase factor in a cylindrical coordinate $(\rho,\phi,z)$ by omitting the higher orders than the second order of $x/z, y/z, a/z, \lambda/z$. Moreover, we perform the Jacobi-Anger expansion to expand the electromagnetic field by the superposition of the vortex field which propagates along the $z$ axis.
\begin{align}
e^{ik|\xi_j|}\approx& \exp\left[ikz\left(1+\frac{\rho^2+a^2}{2z^2}\right)\right]\exp\left[\frac{ik\rho a}{z}\cos(\phi-\phi_j-\pi)\right]\nonumber\\
=&e^{ik\psi(z)}\sum_{m}(-i)^mJ_m\left(\frac{k\rho a}{z}\right)e^{im(\phi-\varphi_j)}
\end{align} 
where $\psi(z)\equiv z\left(1+\frac{\rho^2+a^2}{2z^2}\right)$ is the phase factor which is independent on the $j$ and the azimuthal angle of $\phi$. Secondly, we approximately evaluate the first vector terms of electric and magnetic field. In this approximation, we also omit the terms which are higher than the second order of $x/z,y/z,a/z,\lambda/z$.
\begin{align}
\left(\frac{\pmb{ \xi}_j}{\left|\pmb{ \xi}_j\right|}\times{\bf p}\right)\times\frac{\pmb{ \xi}_j}{\left|\pmb{ \xi}_j\right|^2}
\approx&\frac{pe^{iq\varphi_j}}{z^3}\left( \begin{array}{c}s_x z^2-\frac{3}{2}s_x \xi_x^2-\frac{1}{2}s_x \xi_y^2-s_y\xi_x\xi_y\\
s_y z^2 -\frac{3}{2}s_y \xi_y^2-\frac{1}{2}s_y \xi_x^2 -s_x \xi_x \xi_y\\
-z(s_x \xi_x+s_y \xi_y)
\end{array}\right)\equiv \frac{\pmb{\Lambda}_j}{z}e^{iq\varphi_j}\\
\left(\frac{\pmb{ \xi}_j}{\left|\pmb{ \xi}_j\right|}\times{\bf p}\right)\frac{1}{\left|\pmb{ \xi}_j\right|} 
\approx&\frac{pe^{iq\varphi_j}}{z^3}\left( \begin{array}{c}-s_y z^2 +s_y(\xi_x^2+\xi_y^2)\\ s_x z^2-s_x(\xi_x^2+\xi_y^2)\\-s_x\xi_yz+s_y\xi_xz\end{array}\right)\equiv \frac{\pmb{\Pi}_j}{z} e^{iq\varphi_j}
\end{align}
The terms of $\pmb{\Lambda}_j$ and $\pmb{\Pi}_j$ imply a presence of the spin-orbit interaction of light due to the non-paraxial emission of the electromagnetic fields. By calculating the sum of $\pmb{\Lambda}_j e^{i(q-m)\varphi_j}$ and $\pmb{\Pi}_j e^{i(q-m)\varphi_j}$ (the detail calculation is shown in the Appendix A), the far-field and the near-axis solution of the electromagnetic field is obtained as the following;
\begin{align}
{\bf E}({\bf r})\approx& \frac{k^2}{4\pi\varepsilon_0 z}e^{ik\psi(z)}\sum_Q\left( \begin{array}{c} s_x F^0_Q({\bf r})+F^{SOI}_Q({\bf r}) \\ s_y F^0_Q({\bf r})-i F^{SOI}_Q({\bf r}) \\ G^{SOI,+}_Q({\bf r})\end{array}\right)\\
{\bf H}({\bf r})\approx&\frac{ck}{4\pi z}e^{ik\psi(z)}\sum_Q\left( \begin{array}{c} -s_y F^0_Q({\bf r})\\ s_x F^0_Q({\bf r})\\ iG^{SOI,-}_Q({\bf r})\end{array}\right)
\end{align}
where
\begin{align}
F^0_Q({\bf r})=&\left[\left(1-\frac{\rho^2+a^2}{z^2}\right)f_Q+\frac{\rho a}{z^2}\left(f_{Q+1}+f_{Q-1})\right)\right]e^{iQ\phi},\\
F^{SOI}_Q({\bf r})=&s_-\left[\frac{\rho a}{2z^2}f_{Q+1}-\frac{\rho^2}{4z^2}f_Q-\frac{a^2}{4z^2}f_{Q+2}\right]e^{i(Q+2)\phi}\nonumber\\
+& s_+\left[\frac{\rho a}{2z^2}f_{Q-1}-\frac{\rho^2}{4z^2}f_Q-\frac{a^2}{4z^2}f_{Q-2}\right]e^{i(Q-2)\phi},\\
G_Q^{SOI,\pm}({\bf r})=&s_-\left(\frac{\rho}{2z}f_Q-\frac{a}{2z}f_{Q+1}\right)e^{i(Q+1)\phi}\pm s_+\left(\frac{\rho}{2z}f_Q-\frac{a}{2z}f_{Q-1}\right)e^{i(Q-1)\phi},
\end{align}
$Q\equiv q-nN$ is an integer defined by the system parameters of $N, q$ and an arbitrary integer of $n$, the summation of $Q$ is taken from $n=-\infty$ to $n=\infty$, $f_Q\equiv f_Q(\rho,z)=(-i)^QJ_{Q}\left(\frac{k\rho a}{2z}\right)$ and $s_\pm=s_x\pm is_y$. The term of "SOI" stands for the spin-orbit interaction because the $F^{SOI}_Q({\bf r})$ and $G_Q^{SOI,\pm}({\bf r})$ include the multiplication terms including the opposite sign of the spin term $s_\pm$ and the orbit term $e^{i(Q\mp 1)\phi}$ or $e^{i(Q\mp 2)\phi}$. Using the polynomial expansion of the Bessel function, the lowest order of $k\rho a/z$ is determined by the order of the absolute value of $Q$.
\begin{align}
J_Q\left(\frac{k\rho a}{z}\right)=& (-1)^{\frac{sgn(Q)-1}{2}|Q|}\sum_{t=0}^\infty\frac{(-1)^t}{t!(t+|Q|)!}\left(\frac{k\rho a}{2z}\right)^{2t+|Q|}\approx   (-1)^{\frac{sgn(Q)-1}{2}|Q|}\frac{1}{|Q|!}\left(\frac{k\rho a}{2z}\right)^{|Q|}
\end{align}
where $sgn(x)$ is the sign function. Here, we define an integer $Q_{\mathrm{min}}$ such that $|Q_{\mathrm{min}}|$ becomes the minimum integer of $Q(n)$. If the minimum absolute value of $Q$ does not uniquely defined such that $Q(n_0)=-Q_{\mathrm{min}}$ and $Q(n_1)=Q_{\mathrm{min}}$, there are no optical vortices due to the superposition of the clockwise and the anti-clockwise vortices. For example, this situation appears in the case of $Q_{\mathrm{min}}=0$. On the other hand, in the case where the minimum absolute value is uniquely defined by $n$, each term of the electromagnetic field are simply expressed as a form including optical vortices with several terms for each $Q_{\mathrm{min}}$.
\begin{align}
\sum_Q F^0_Q({\bf r})\approx&\frac{(-i)^{|Q_{\mathrm{min}}|}}{|Q_{\mathrm{min}}|!}(-1)^{\frac{sgn(Q_{\mathrm{min}})-1}{2}|Q_{\mathrm{min}}|} \left(\frac{k\rho a}{2z}\right)^{Q_{\mathrm{min}}} e^{iQ_{\mathrm{min}}\phi}\\
\sum_Q F^{SOI}_Q({\bf r})\approx&\left\{ \begin{array}{cc}0 & (Q_{\mathrm{min}}=\pm 1)\\
-s_+ \frac{(-i)^{Q_{\mathrm{min}}-2}}{(Q_{\mathrm{min}}-2)!}\left(\frac{k\rho a}{2z}\right)^{Q_{\mathrm{min}}-2}\frac{a^2}{4z^2}e^{i(Q_{\mathrm{min}}-2)\phi}& (Q_{\mathrm{min}}\geq 2)\\
-(-1)^{|Q_{\mathrm{min}}+2|}s_- \frac{i^{|Q_{\mathrm{min}}+2|}}{|Q_{\mathrm{min}}+2|!}\left(\frac{k\rho a}{2z}\right)^{|Q_{\mathrm{min}}+2|}\frac{a^2}{4z^2}e^{i(Q_{\mathrm{min}}+2)\phi}& (Q_{\mathrm{min}}\leq -2)\end{array}\right.\\
\sum_Q G_Q^{SOI,\pm}({\bf r})\approx&\left\{ \begin{array}{cc}\mp s_+ \frac{(-i)^{Q_{\mathrm{min}}-1}}{(Q_{\mathrm{min}}-1)!}\left(\frac{k\rho a}{2z}\right)^{Q_{\mathrm{min}}-1}\frac{a}{2z}e^{i(Q_{\mathrm{min}}-1)\phi}& (Q_{\mathrm{min}}\geq 1)\\
(-1)^{|Q_{\mathrm{min}}+1|}s_-  \frac{i^{|Q_{\mathrm{min}}+1|}}{|Q_{\mathrm{min}}+1|!}\left(\frac{k\rho a}{2z}\right)^{|Q_{\mathrm{min}}+1|}\frac{a}{2z}e^{i(Q_{\mathrm{min}}+1)\phi}& (Q_{\mathrm{min}}\leq -1)
\end{array} \right.
\end{align}
The analytical solution of the electromagnetic field expressed by the vortex fields in the Eq. (12)--(14) is one of our main results. In the far-field and the near-axis regime, we obtain a certain relationship between the source conditions and the topological charge of the optical vortices. Figure 2 show the value of $Q_{\mathrm{min}}$ with respect to the system parameters $(N,q)$ [Fig. 2(a)], and several phase profiles calculated from the exact solutions in Eq. (1) and (2) fixing the collective polarization to $(s_x=1,s_y=0)$ [Fig. 2(b)--2(d)]. $Q_{\mathrm{min}}$ obviously corresponds to the topological charge calculated from the exact solution, and becomes a good index for characterizing the dipole array as an optical vortex emitter corresponds to the topological charge calculated from the exact solution.
\begin{figure}[htbp]
\centering\includegraphics[width=13cm]{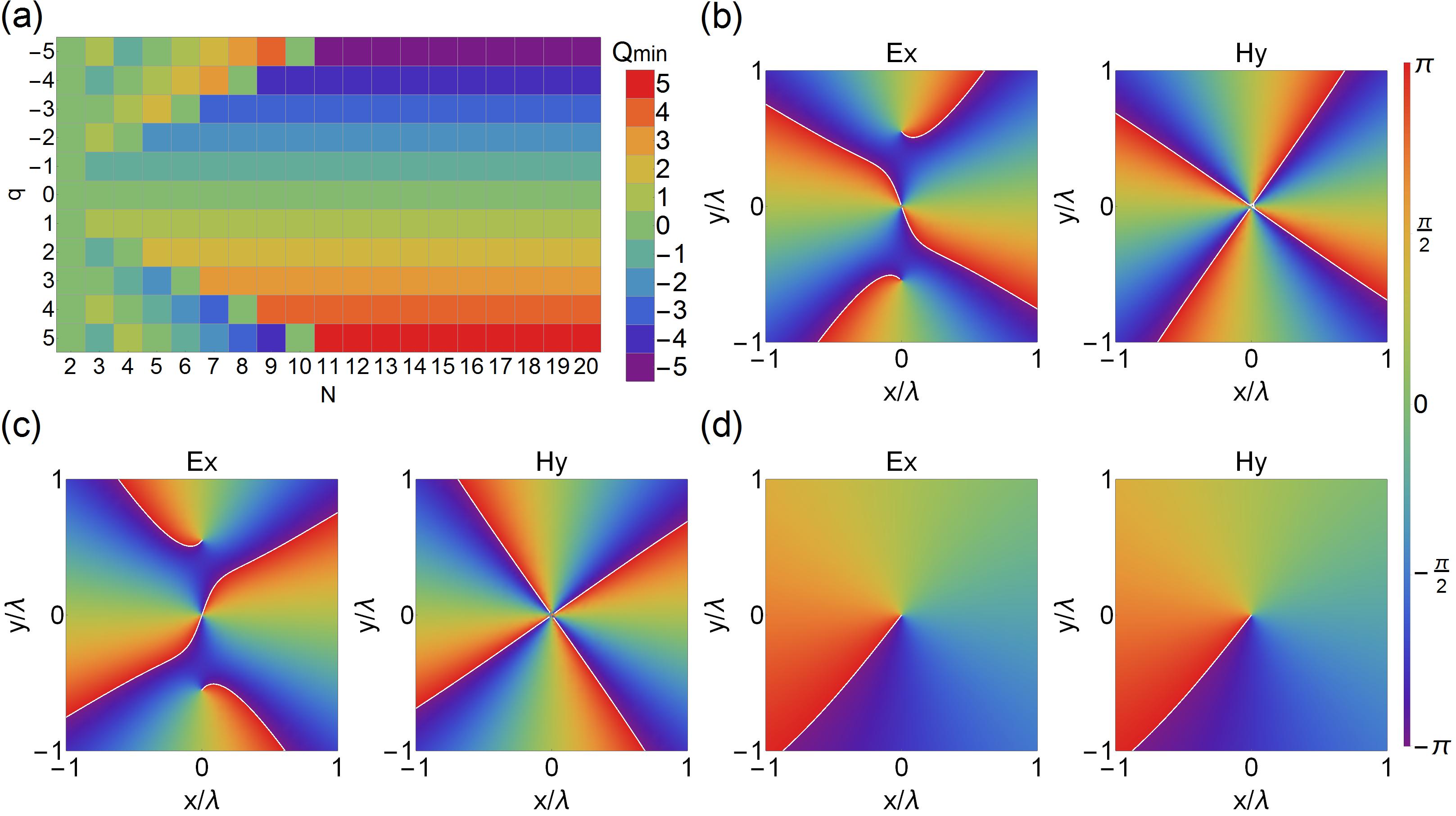}
\vspace{10pt}
\caption{The topological charge of the optical vortex with respect to the source conditions with the collective polarization of $(s_x=1,s_y=0)$. (a) The relationship between $Q_{\mathrm{min}}$ and $(N,q)$. The phase profiles of the electric and the magnetic fields are shown in (b) $\sim$ (d) with the source conditions of $(N,q)=(9,5)$, $(N,q)=(9,4)$ and $(N,q)=(9,1)$ respectively}
\end{figure}

\subsection{Intrinsic split of singular points in electric field with higher order topological charge} 
\begin{figure}[htbp]
\centering\includegraphics[width=13cm]{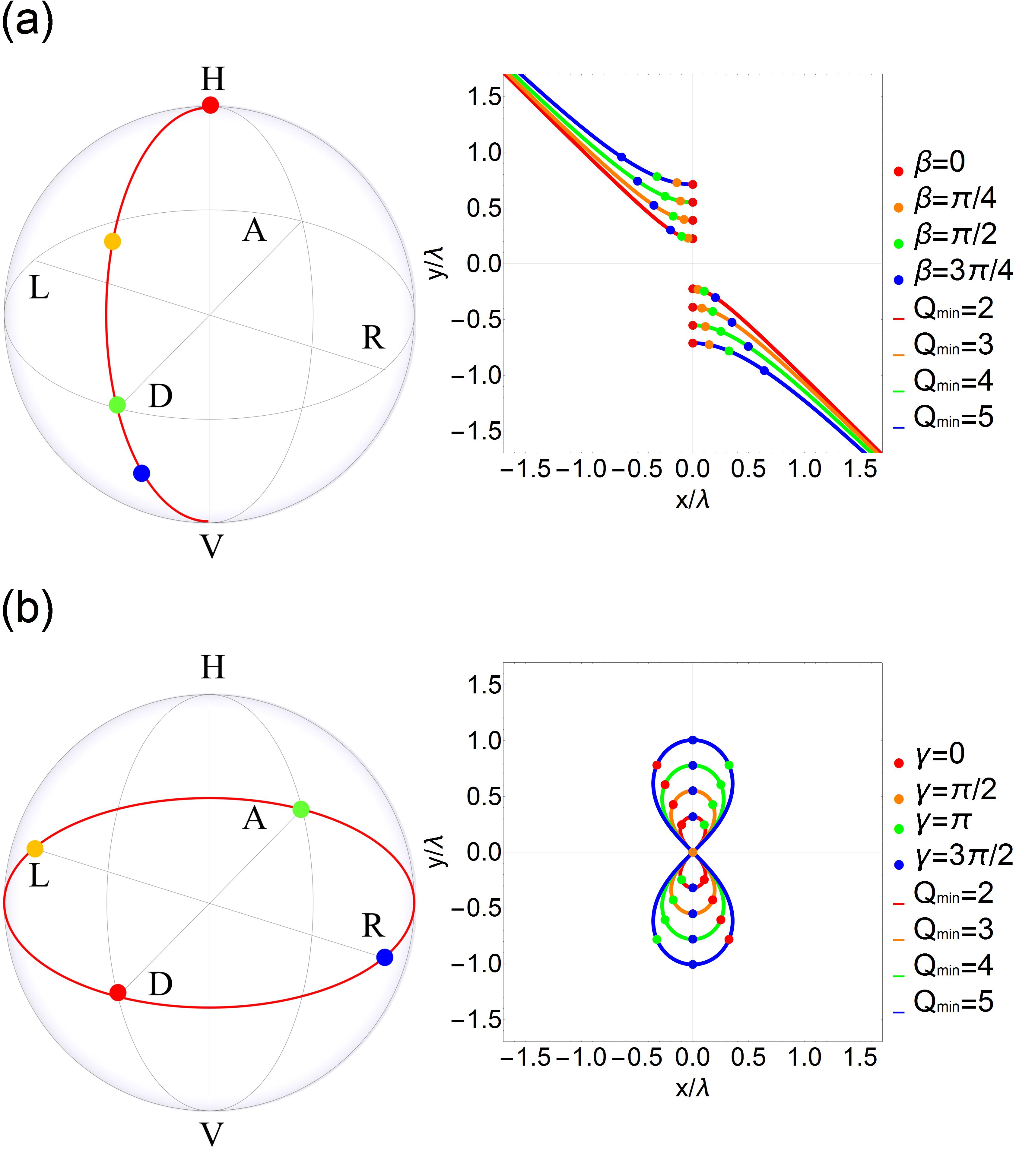}
\vspace{10pt}
\caption{The correspondence between the coordinates of the singular points (right) and the collective polarizations on the Poincare sphere (left). The dots of same color in both of the left and the right fiure shows the same angle of $\beta$ and $\gamma$. The colered lines in the right figure show the trajectory in the same $Q_{\mathrm{min}}$. The characters in the left figure indicate the collective polarization states. H: horizontal polarization, V: vertical polarization, D: diagonal polarization, A: anti-diagonal polarization, L: left circular polarization and R: right circular polarization.}
\end{figure}
In the case where $|Q_{\mathrm{min}}|\geq 2$, the transverse components of electric fields are represented by the sum of the two terms. This implies that the singular points of the transverse components are not located at the center of $xy$ plane. For simplicity, we consider only the case of $Q_{\mathrm{min}}\geq 2$ in the following discussion. The transverse components of the electric field are given by
\begin{align}
E_x({\bf r})\approx&-\frac{k^2}{4\pi\varepsilon_0 z}\frac{(-ik\rho)^{Q_{\mathrm{min}}-2}}{(Q_{\mathrm{min}}-2)!}\left(\frac{a}{2z}\right)^{Q_{\mathrm{min}}}e^{i(Q_{\mathrm{min}}-2)\phi}\left[\frac{s_x}{Q_{\mathrm{min}}(Q_{\mathrm{min}}-1)}(k\rho)^2e^{i2\phi}+s_+\right]\\
E_y({\bf r})\approx&\frac{k^2}{4\pi\varepsilon_0 z}\frac{(-ik\rho)^{Q_{\mathrm{min}}-2}}{(Q_{\mathrm{min}}-2)!}\left(\frac{a}{2z}\right)^{Q_{\mathrm{min}}}e^{i(Q_{\mathrm{min}}-2)\phi}\left[\frac{s_y}{Q_{\mathrm{min}}(Q_{\mathrm{min}}-1)}(k\rho)^2e^{i2\phi}+is_+\right].
\end{align}
The singular points with the topological charge $Q_{\mathrm{min}}-2$ are located at the center of $xy$ plane. In addition, there are non-trivial singular points hidden in the sum of the final terms. In order to expand a precise calculation for the coordinate of all singular points, we express the compolex polarization paramter as a state vector on the Poincare sphere,$s_x=\cos\frac{\beta}{2}$, $s_y=\sin\frac{\beta}{2}e^{i\gamma}$ where each angle $\beta$ and $\gamma$ are bounded as $0\leq \beta\leq \pi$ and $0\leq \gamma \leq 2\pi$ respectively. There are two non-trivial singular points with the topological charge one in the coordinate of $\pmb{\zeta}^i=(\zeta^i_x,\zeta^i_y)$ where $i\in\{1,2\}$ shows the label of the two singular points (the detail calculation is shown in the Appendix B).
\begin{align}
{\pmb \zeta}^1(\beta,\gamma)=&\left\{\begin{array}{ll} (\zeta_+(\beta,\gamma),-\zeta_-(\beta,\gamma))&(0\leq \gamma\leq \frac{\pi}{2},\frac{3\pi}{2}\leq \gamma \leq 2\pi)\\
 (\zeta_+(\beta,\gamma),\zeta_-(\beta,\gamma))&(\frac{\pi}{2} \leq \gamma \leq 2\pi)
\end{array}\right.\\
{\pmb \zeta}^2(\beta,\gamma)=&\left\{\begin{array}{ll} (-\zeta_+(\beta,\gamma),\zeta_-(\beta,\gamma))&(0\leq \gamma\leq \frac{\pi}{2},\frac{3\pi}{2}\leq \gamma \leq 2\pi)\\
 (-\zeta_+(\beta,\gamma),-\zeta_-(\beta,\gamma))&(\frac{\pi}{2} \leq \gamma \leq 2\pi)
\end{array}\right.\\
\zeta_\pm(\beta,\gamma)\equiv&\pm\frac{\lambda\sqrt{Q_{\mathrm{min}}(Q_{\mathrm{min}}-1)}}{2\pi\sqrt{2}}\sqrt{\frac{\sqrt{1-\sin\beta\sin\gamma}}{\cos\frac{\beta}{2}}\pm\left(\tan\frac{\beta}{2}\sin\gamma -1\right)}
\end{align}
Because the coordinates of the singular point are independent on $z$ coordinate, all singular points propagate in parallel to the $z$ axis in the far-field regime. Equation (17)--(19) are completely characterized the intrinsic split of the singular points in the electric field of $E_x$ with respect to the source condition of the emitter. If the electric dipoles are horizontally polarized ($\beta=0$ and $\gamma=0$), the singular points in the electric field of $E_x$ are vertically split with the distance of the singular points simply expressed by
\begin{align}
D=\frac{\lambda\sqrt{Q_{\mathrm{min}}(Q_{\mathrm{min}}-1)}}{2\pi}.
\end{align}
On the other hand, if the electric dipoles are left (right) circularly polarized in the case of the positive (negative) $Q_{\mathrm{min}}$, the singlar points in the electric field are not split and are located at the center of the $xy$ plane with the topological charge $Q_{\mathrm{min}}$. We can make a single optical vortex with the topological charge $Q_{\mathrm{min}}$ propagating on the $z$ axis. Several correspondence between the transverse coordinate of the singular points and the collective polarization of the electric dipoles on the Poincare sphere are shown in Fig. 3. 

The split of the higher-order optical vortex has been discussed in the holograhic method \cite{basistiy1993optics,ricci2012instability}. In general, the higher-order vortex is easily split into several vortices with the topological charge one under the practical fluctuation and incompleteness. We emphasize that if the perfect system removing any practical problems is assumed for generation of the optical vortices, nevertheless, there exists intrinsic split of the singular points in the electric field in the dipole array due to the spin-orbit interaction caused by the radiation patterns of each electric dipole.

\section{Numerical simulation}
\begin{figure}[htbp]
\centering\includegraphics[width=13cm]{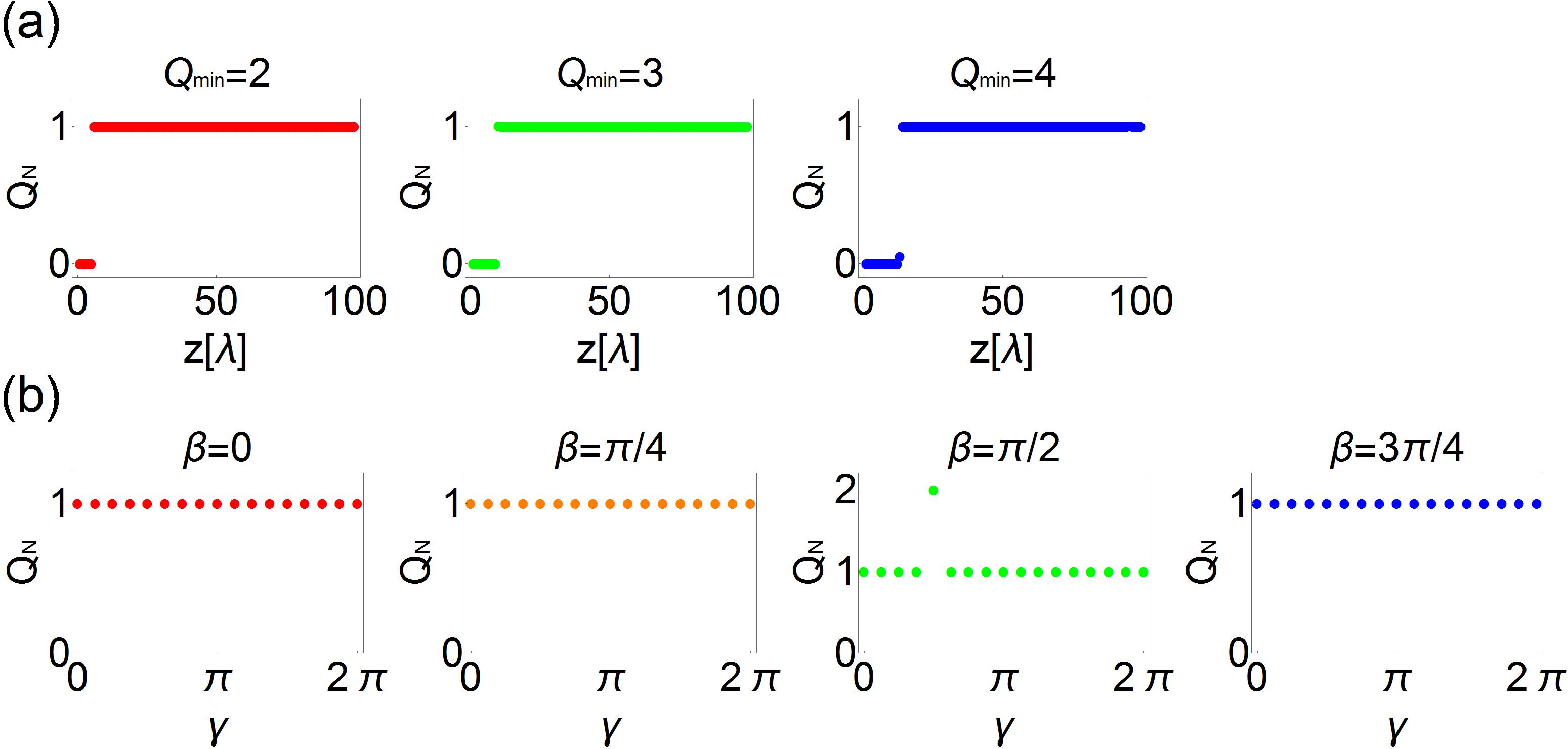}
\vspace{10pt}
\caption{The results of the numerical calculations (a) the topological charges around the coordinate of ${\pmb \zeta}^1(0,0)$ for each $z$. (b) the topological charges around the coordinate of ${\pmb\zeta}^1(\beta,\gamma)$ at $z=100\lambda$ for each $\beta$ and $\gamma$.}
\end{figure}
To verify the consistency of our analytical approach with several approximations, we numerically calculate a topological charge around one of the singular points $\boldsymbol \zeta^1(\beta,\gamma)$ from the exact solution of Eq. (1) and (2). The topological charge that we numerically calculated is defined as
\begin{align}
Q_{N}\equiv &\frac{1}{2\pi}\oint_c \mathrm{d}l \nabla \mathrm{Arg}\left[E_x({\bf r})\right]\nonumber\\
=&\frac{1}{2\pi}\int_{-\pi}^{\pi}\mathrm{d}\phi_1 \mathrm{Arg}\left[E_x(R_N\cos\phi_1,R_N\sin\phi_1,z)\right]  
\end{align}
where $R_N$ is a radius of closed path for integration and $\phi_1$ is the azimuthal angle around ${\pmb \zeta}^1(\beta,\gamma)$. The topological charge depends on the $z$ coordinate when the state of the polarization is fixed to $\beta=0$ and $\gamma=0$ [Fig. 4(a)]. The topological charges for each $Q_{\mathrm{min}}$ reach to one when the $z$ coordinate is over several ten times of the wavelength. This implies that there is a split vortex with a topological charge one in the coordinate of the singular points given by Eq. (17)--(19) around the far-field regime, and the singular points propagate in parallel to the $z$ axis. Figure 4(b) shows the numerical topological charges under $Q_{\mathrm{min}}=2$. At all of the points without the case where $\beta=\gamma=\pi/2$, the singular point with the topological charge one appears. This implies that the singular points split into two, and one of them propagates along $z$ axis with the coordinate of ${\pmb \zeta}^1(\beta,\gamma)$. On the other hand, there is an optical vortex with the topological charge two which has no splitting at the center of the $xy$ plane in the case of $\beta=\gamma=\pi/2$ because the coordinate of singular points degenerate into ${\pmb\zeta}^1(\beta,\gamma)=(0,0)$. All numerical result calculated from the exact solutions is supported by our analytical approach with the far-field and the near-axis approximations.

\section{Discussion}
We mainly focused on the phase distribution of the electromagnetic field to characterize the structures of the optical vortices generated by the electric dipole array. On the other hand, the energy distribution is also important to discuss the function of the system. Figure 5 shows the energy density in the $zx$ plane calculated from the exact solutions for each source condition $Q_{\mathrm{min}}$ with the polarization condition of $(s_x,s_y)=(1,0)$. The larger is the area where the energy becomes weak around the $z$ axis, the larger is the topological charge of the optical vortex. It stems from the terms of amplitude including the Bessel function with the order of its topological charge.
\begin{figure}[h]
\centering\includegraphics[width=13cm]{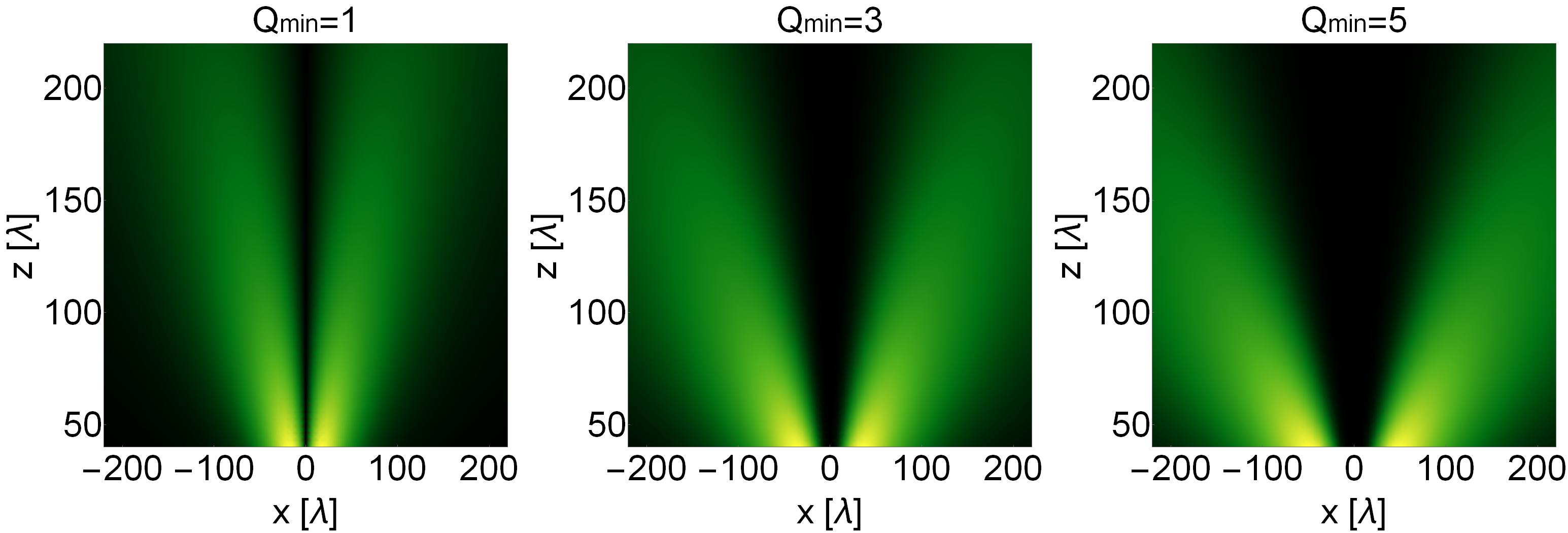}
\vspace{10pt}
\caption{Amplitude profiles of the optical vortex in each topological charge}
\end{figure}

An electric dipole is the most fundamental element for emitting electromagnetic field to the far-field with a wide range of wavelengths. For the implementation of our optical vortex generator, it is necessary to fix the position of each dipole on a circle and to control each phase. It is possible to easily implement our system in radiowave or microwave regions because the phase of the electric dipoles can be directly controlled with conventional systems. In addition, the intrinsic split of the singular points in the electric field can be experimentally detectable due to the long wavelengths. Our system has a possibility to switch the topological charges only by controlling the phase distributions of the electric dipoles $q$ without the number of the electric dipoles $N$.

\section{Conclusion}
In this paper, we discuss the propagation properties of the optical vortices emitted from a collectively polarized dipole array. Under the far-field and the near-axis approximations, we clarify the following two properties: (1) there is a certain relationship between the source conditions and the topological charge of the optical vortices, (2) There exists an intrinsic split of the singular points in the electric field due to the spin-orbit interaction of light between the collective electric dipole moments and the intrinsic radiation patterns of each emitter without any split of the singular point in the magnetic field. These analytical results give us a certain perspective for controlling the topological charge of optical vortices.

\section*{Acknowledgments}
This work was supported by Program for Leading Graduate Schools: "Interactive Materials Science Cadet Program".

\appendix
\section{Derivation of the far-field solution}
In order to derive the far-field solution by decomposing the optical vortices which propagates on the $z$ axis, we start from the electric and the magnetic field indicated by the superposition of the field emitted from each dipole
\begin{align}
{\bf E}({\bf r})&\approx \frac{k^2}{4\pi\varepsilon_0 z}\sum_{m=-\infty}^\infty (-i)^m J_{m}\left(\frac{k\rho a}{z}\right)e^{im\phi}\sum_{j=1}^N \pmb{ \Lambda}_j e^{i(q-m)\varphi_j}\\
{\bf H}({\bf r})&\approx \frac{ck}{4\pi z}\sum_{m=-\infty}^\infty (-i)^m J_{m}\left(\frac{k\rho a}{z}\right)e^{im\phi}\sum_{j=1}^N \pmb{ \Pi}_j e^{i(q-m)\varphi_j},
\end{align}
where $\pmb{\Lambda}_j$ and $\pmb{\Pi}_j$ are the spin-orbit interaction terms defined in Eq. (4) and (5) respectively. We obtain the selection rules for the topological charges given by the Dirac's delta via the direct calculation of the sum of $\pmb{\Lambda}_j$ and $\pmb{\Pi}_j$.

\begin{align}
\left[\pmb{ \Lambda_j}\right]_x=&s_x\Biggl(1-\frac{3\rho^2}{2z^2}\cos^2\phi-\frac{3a^2}{2z^2}\cos^2\varphi_j +\frac{3\rho a}{z^2} \cos\phi\cos\varphi_j\nonumber\\
&-\frac{\rho^2}{2z^2}\sin^2\phi-\frac{a^2}{2z^2}\sin^2\varphi_j+\frac{\rho a}{z^2} \sin\phi\sin\varphi_j\Biggr)\nonumber\\
&-s_y\left\{\frac{\rho^2}{z^2}\cos\phi\sin\phi+\frac{a^2}{z^2}\cos\varphi_j\sin\varphi_j-\frac{\rho a}{z^2}\left(\cos\phi\sin\varphi_j+\cos\varphi_j\sin\phi\right)\right\}\nonumber\\
=&s_x\Biggl(1-\frac{\rho^2+a^2}{z^2}-\frac{\rho^2}{2z^2}\cos2\phi-\frac{a^2}{2z^2}\cos2\varphi_j\nonumber\\s
&+\frac{3\rho a}{z^2} \cos\phi\cos\varphi_j+\frac{\rho a}{z^2} \sin\phi\sin\varphi_j\Biggr)\nonumber\\
&-s_y\left(\frac{\rho^2}{2z^2}\sin2\phi+\frac{a^2}{2z^2}\sin2\varphi_j-\frac{\rho a}{z^2}\sin\left(\phi+\varphi_j\right)\right)\\
%% x component
\sum_{j=1}^N \left[\pmb{ \Lambda_j}\right]_x e^{i(q-m)\varphi_j}=&\sum_{n=-\infty}^\infty s_x\Biggl[\left(1-\frac{\rho^2+a^2}{z^2}\right)\delta_{q-m,nN}-\frac{\rho^2}{4z^2}\left(e^{i2\phi}+e^{-i2\phi}\right)\delta_{q-m,nN}\nonumber\\
&-\frac{a^2}{4z^2}\left(\delta_{q-m+2,nN}+\delta_{q-m-2,nN}\right)\nonumber\\
&+\frac{3\rho a}{4z^2} \left(e^{i\phi}+e^{-i\phi}\right)\left(\delta_{q-m+1,nN}+\delta_{q-m-1,nN}\right)\nonumber\\
&-\frac{\rho a}{4z^2} \left(e^{i\phi}-e^{-i\phi}\right)\left(\delta_{q-m+1,nN}-\delta_{q-m-1,nN}\right)\Biggr]\nonumber\\
&-\sum_{n=-\infty}^\infty s_y\Biggl[\frac{\rho^2}{i4z^2}\left(e^{i2\phi}-e^{-i2\phi}\right)\delta_{q-m,nN}+\frac{a^2}{i4z^2}\left(\delta_{q-m+2,nN}-\delta_{q-m-2,nN}\right)\nonumber\\
&-\frac{\rho a}{i2z^2}\left(e^{i\phi}\delta_{q-m+1,nN}-e^{-i\phi}\delta_{q-m-1,nN}\right)\Biggr]\\
%% y component
\sum_{j=1}^N \left[\pmb{ \Lambda_j}\right]_y e^{i(q-m)\varphi_j}=&\sum_{n=-\infty}^\infty s_y\Biggl[\left(1-\frac{\rho^2+a^2}{z^2}\right)\delta_{q-m,nN}+\frac{\rho^2}{4z^2}\left(e^{i2\phi}+e^{-i2\phi}\right)\delta_{q-m,nN}\nonumber\\
&+\frac{a^2}{4z^2}\left(\delta_{q-m+2,nN}+\delta_{q-m-2,nN}\right)\nonumber\\
&-\frac{3\rho a}{4z^2} \left(e^{i\phi}-e^{-i\phi}\right)\left(\delta_{q-m+1,nN}-\delta_{q-m-1,nN}\right)\nonumber\\
&+\frac{\rho a}{4z^2} \left(e^{i\phi}+e^{-i\phi}\right)\left(\delta_{q-m+1,nN}+\delta_{q-m-1,nN}\right)\Biggr]\nonumber\\
&-\sum_{n=-\infty}^\infty s_x\Biggl[\frac{\rho^2}{i4z^2}\left(e^{i2\phi}-e^{-i2\phi}\right)\delta_{q-m,nN}+\frac{a^2}{i4z^2}\left(\delta_{q-m+2,nN}-\delta_{q-m-2,nN}\right)\nonumber\\
&-\frac{\rho a}{i2z^2}\left(e^{i\phi}\delta_{q-m+1,nN}-e^{-i\phi}\delta_{q-m-1,nN}\right)\Biggr]\\
%% z component
\sum_{j=1}^N \left[\pmb{ \Lambda_j}\right]_z e^{i(q-m)\varphi_j}=&-\sum_{n=-\infty}^\infty \Biggl[s_x\left\{\frac{\rho}{2z}\left(e^{i\phi}+e^{-i\phi}\right)\delta_{q-m,nN}-\frac{a}{2z}\left(\delta_{q-m+1,nN}+\delta_{q-m-1,nN}\right)\right\}\nonumber\\
&+s_y\left\{\frac{\rho}{i2z}\left(e^{i\phi}-e^{-i\phi}\right)\delta_{q-m,nN}-\frac{a}{i2z}\left(\delta_{q-m+1,nN}-\delta_{q-m-1,nN}\right)\right\}\Biggr]
\end{align}
Finally, all component of the electric field is calculated as 
\begin{align}
E_x({\bf r})\approx& \frac{k^2}{4\pi\varepsilon_0 z}\sum_{Q} s_x\left(f_Q-\frac{\rho^2+a^2}{z^2}f_Q+\frac{\rho a}{z^2}\left(f_{Q+1}+f_{Q-1})\right)\right)e^{iQ\phi}\nonumber\\
+&\sum_{Q} s_-\left[\frac{\rho a}{2z^2}f_{Q+1}-\frac{\rho^2}{4z^2}f_Q-\frac{a^2}{4z^2}f_{Q+2}\right]e^{i(Q+2)\phi}\nonumber\\
+&\sum_{Q} s_+\left[\frac{\rho a}{2z^2}f_{Q-1}-\frac{\rho^2}{4z^2}f_Q-\frac{a^2}{4z^2}f_{Q-2}\right]e^{i(Q-2)\phi},\\
%%y component
E_y({\bf r})\approx& \frac{k^2}{4\pi\varepsilon_0 z}\sum_{Q} s_y\left(f_Q-\frac{\rho^2+a^2}{z^2}f_Q+\frac{\rho a}{z^2}\left(f_{Q+1}+f_{Q-1})\right)\right)e^{iQ\phi}\nonumber\\
-i&\sum_{Q} s_-\left[\frac{\rho a}{2z^2}f_{Q+1}-\frac{\rho^2}{4z^2}f_Q-\frac{a^2}{4z^2}f_{Q+2}\right]e^{i(Q+2)\phi}\nonumber\\
-i&\sum_{Q} s_+\left[\frac{\rho a}{2z^2}f_{Q-1}-\frac{\rho^2}{4z^2}f_Q-\frac{a^2}{4z^2}f_{Q-2}\right]e^{i(Q-2)\phi},\\
%%z component
E_z({\bf r})\approx& \frac{k^2}{4\pi\varepsilon_0 z}\sum_{Q}\left[ s_-\left(\frac{\rho}{2z}f_Q-\frac{a}{2z}f_{Q+1}\right)e^{i(Q+1)\phi}+s_+\left(\frac{\rho}{2z}f_Q-\frac{a}{2z}f_{Q-1}\right)e^{i(Q-1)\phi}\right],
\end{align}
where $f_n\equiv f_n(x)\equiv (-i)^n J_n(x)$. As a same manner, the magnetic field is calculated as
\begin{align}
\Pi_j=&\frac{1}{z^2}\left( \begin{array}{c}-s_y z^2 +s_y(\xi_x^2+\xi_y^2)\\ s_x z^2-s_x(\xi_x^2+\xi_y^2)\\-s_x\xi_yz+s_y\xi_xz\end{array}\right),\\
%% x component
\sum_{j=1}^N \left[\pmb{ \Pi_j}\right]_x e^{i(q-m)\varphi_j}=&-s_y\sum_{n}\Biggl[\left(1-\frac{\rho^2+a^2}{z^2}\right)\delta_{q-m,nN}\nonumber\\
&+\frac{\rho a}{z^2}\left(e^{i\phi}\delta_{q-m-1,nN}+e^{-i\phi}\delta_{q-m+1,nN}\right)\Biggr],\\
%% y component
\sum_{j=1}^N \left[\pmb{ \Pi_j}\right]_y e^{i(q-m)\varphi_j}=&s_x\sum_{n}\Biggl[\left(1-\frac{\rho^2+a^2}{z^2}\right)\delta_{q-m,nN}\nonumber\\
&+\frac{\rho a}{z^2}\left(e^{i\phi}\delta_{q-m-1,nN}+e^{-i\phi}\delta_{q-m+1,nN}\right)\Biggr],\\
%% z component
\sum_{j=1}^N \left[\pmb{ \Pi_j}\right]_z e^{i(q-m)\varphi_j}=&is_-\sum_{n}\left(\frac{\rho}{2z}e^{i\phi}\delta_{q-m,nN}-\frac{a}{2z}\delta_{q-m+1,nN}\right)\nonumber\\
&-is_+\sum_{n}\left(\frac{\rho}{2z}e^{-i\phi}\delta_{q-m,nN}-\frac{a}{2z}\delta_{q-m-1,nN}\right),\\
%% x component
H_x({\bf r})=&-\frac{ck}{4\pi z}s_ye^{iQ\phi}\sum_{n}\left[\left(1-\frac{\rho^2+a^2}{z^2}\right)f_{Q}+\frac{\rho a}{z^2}\left(f_{Q-1}+f_{Q+1}\right)\right],\\
%% y component
H_y({\bf r})=&\frac{ck}{4\pi z}s_xe^{iQ\phi}\sum_{n}\left[\left(1-\frac{\rho^2+a^2}{z^2}\right)f_{Q}+\frac{\rho a}{z^2}\left(f_{Q-1}+f_{Q+1}\right)\right],\\
%% z component
H_z({\bf r})=&\frac{ick}{4\pi z}\Biggl[s_-e^{i(Q+1)\phi}\sum_{n}\left(\frac{\rho}{2z}f_Q-f_{Q+1}\right)\nonumber\\
&-s_+e^{i(Q-1)\phi}\sum_{n}\left(\frac{\rho}{2z}f_Q-\frac{a}{2z}f_{Q-1}\right)\Biggr].
\end{align}

\section{Derivation of the singular points in $E_x$}
In the case $Q_{\mathrm{min}}\geq 2$, the $E_x$ given in Eq. (15) includes the non-trivial vortices. In order to find the coordinate of the singular points of the non-trivial vortices, we derive the product form of $e^{i\phi_1} e^{i\phi_2}$ from the final terms of Eq. (15) where $\phi_1$ and $\phi_2$ are defined as new azimuthal angles in cylindrical coordinates achieved by translating the $z$ axis into the radial directions respectively. At first, we define $A=\frac{k^2}{Q_{\mathrm{min}}(Q_{\mathrm{min}}-1)}$, and expand to the Cartesian coordinate to find the transformation factor $\zeta_x$ and $\zeta_y$.
\begin{align}
As_x\rho^2e^{i2\phi}+s_+=&\left[A\cos\frac{\beta}{2}(x^2-y^2+i2xy)+\cos\frac{\beta}{2}+i\sin\frac{\beta}{2}\cos\gamma-\sin\frac{\beta}{2}\sin\gamma\right]\nonumber\\
=&\left[A\cos\frac{\beta}{2}(x^2-y^2)+\cos\frac{\beta}{2}-\sin\frac{\beta}{2}\sin\gamma+i\left(2Axy\cos\frac{\beta}{2}+\sin\frac{\beta}{2}\sin\gamma\right)\right]
\end{align}
The phase factor $\phi'$ is calculated as 
\begin{align}
\phi'=&\tan^{-1}\left[\frac{\mathrm{Im}\left[G\right]}{\mathrm{Re}\left[G\right]}\right]=\tan^{-1}\left[\frac{2Axy\cos\frac{\beta}{2}+\sin\frac{\beta}{2}\cos\gamma}{A\cos\frac{\beta}{2}(x^2-y^2)+\cos\frac{\beta}{2}-\sin\frac{\beta}{2}\sin\gamma}\right]\nonumber\\
=&\tan^{-1}\left[\frac{2xy+\frac{1}{A}\tan\frac{\beta}{2}\cos\gamma}{(x^2-y^2)+\frac{1}{A}-\frac{1}{A}\tan\frac{\beta}{2}\sin\gamma}\right]\nonumber.
\end{align}
According to a property of the inverse trigonometric function, the phase factor is separated as a form of sum.
\begin{align}
\tan^{-1}\left[\frac{\frac{y-\zeta_y}{x-\zeta_x}+\frac{y+\zeta_y}{x+\zeta_x}}{1-\frac{y-\zeta_y}{x-\zeta_x}\frac{y+\zeta_y}{x+\zeta_x}}\right]=\tan^{-1}\frac{y+\zeta_y}{x+\zeta_x}+\tan^{-1}\frac{y-\zeta_y}{x-\zeta_x}
\end{align}
Finally, we can find the translation factor $\zeta_x,\zeta_y$ from the following simultaneous equations:
\begin{align}
\zeta_x\zeta_y=&-\frac{1}{2A}\tan\frac{\beta}{2}\cos\gamma,\\
\zeta_y^2-\zeta_x^2=&\frac{1}{A}-\frac{1}{A}\tan\frac{\beta}{2}\sin\gamma.
\end{align}
As a result, we obtain the coordinate of the singular points.
\begin{align}
{\pmb \zeta}^1(\beta,\gamma)=&\left\{\begin{array}{ll} (\zeta_+(\beta,\gamma),-\zeta_-(\beta,\gamma))&(0\leq \gamma\leq \frac{\pi}{2},\frac{3\pi}{2}\leq \gamma \leq 2\pi)\\
 (\zeta_+(\beta,\gamma),\zeta_-(\beta,\gamma))&(\frac{\pi}{2} \leq \gamma \leq 2\pi)
\end{array}\right.\\
{\pmb \zeta}^2(\beta,\gamma)=&\left\{\begin{array}{ll} (-\zeta_+(\beta,\gamma),\zeta_-(\beta,\gamma))&(0\leq \gamma\leq \frac{\pi}{2},\frac{3\pi}{2}\leq \gamma \leq 2\pi)\\
 (-\zeta_+(\beta,\gamma),-\zeta_-(\beta,\gamma))&(\frac{\pi}{2} \leq \gamma \leq 2\pi)
\end{array}\right.\\
\zeta_\pm(\beta,\gamma)\equiv&\pm\frac{\lambda\sqrt{Q_{\mathrm{min}}(Q_{\mathrm{min}}-1)}}{2\pi\sqrt{2}}\sqrt{\frac{\sqrt{1-\sin\beta\sin\gamma}}{\cos\frac{\beta}{2}}\pm\left(\tan\frac{\beta}{2}\sin\gamma -1\right)}
\end{align}

\end{document}